\documentclass[preprint,showpacs,preprintnumbers,amsmath,amssymb]{revtex4}

\usepackage{graphicx}
\usepackage{dcolumn}
\usepackage{bm}

\begin{document}

\preprint{APS/123-QED}

\title{Spin spirals in ordered and disordered solids}

\author{S.~Mankovsky$^1$, G.~H.~Fecher$^2$  and H.~Ebert$^1$}
\affiliation{%
$^1$Department of Chemistry/Phys. Chemistry, LMU Munich,
Butenandtstrasse 11, D-81377 Munich, Germany \\
$^2$ Institut f\"ur Anorg. Chemie und Analyt. Chemie, Johannes Gutenberg-Universit\"at Mainz, D-55099 Mainz, Germany
}%

\date{\today}

\begin{abstract}
 A scheme to calculate the
  electronic structure  of systems having a spiral magnetic structure is
  presented. The approach is based on the KKR  (Korringa-Kohn-Rostoker) Green's function
  formalism which allows in combination with CPA  (Coherent Potential Approximation)
alloy theory to deal
  with chemically disordered materials. It is applied to the 
  magnetic  random
  alloys Fe$_x$Ni$_{1-x}$, Fe$_x$Co$_{1-x}$ and Fe$_x$Mn$_{1-x}$. 
 For these systems the stability of  their magnetic structure was analyzed. For 
Fe$_x$Ni$_{1-x}$  the spin stiffness for was  determined as a
  function of concentration that  was found in satisfying agreement with 
  experiment.
  Performing spin spiral calculations the longitudinal
  momentum-dependent magnetic susceptibility was calculated  for
  pure elemental systems (Cr, Ni) being in non-magnetic state  as  well as for
  random alloys  (Ag$_x$Pt$_{1-x}$). 
  The obtained susceptibility was used to analyze the
stability of  the paramagnetic state of these systems. 
\end{abstract}

\maketitle

\section{ Introduction \label{IN}}

The use of symmetry properties of solids for calculations of their
electronic structure is a very efficient way  to reduce the
computational effort required for the solution of the problem.
In particular, the single-particle electronic states of 
paramagnetic or collinear magnetic infinite solids can be
effectively found by solving the corresponding
Kohn-Sham-Dirac equation  making use of the Bloch theorem.
Dealing with systems exhibiting non-collinear magnetic
structure, the  electronic structure problem  becomes 
much more complicated because of broken symmetry (in general, both
translational and rotational),  leading to an increase of the unit
cell of a system  and a corresponding increase of  the
required computational
effort. 

Sandratskii  introduced an approach  that allows to calculate the electronic structure
 of systems with spiral magnetic structures  in an efficient way \cite{San86,San91}.
 This approach is based on the symmetry properties 
of spin spiral structures   as investigated by
Brinkman and Elliot \cite{BE66a,BE66b} and Herring \cite{Her66} and
allows to
deal with long-period non-collinear magnetic structures avoiding the use
of big unit cells in electronic structure calculations \cite{San98}.
This makes it an efficient tool for the analysis of  the stability of various 
non-collinear magnetic structures with different translation period, as for
example  demonstrated by Mryasov et al.\ \cite{MLSG91} for the investigation of the
magnetic
structure  of  fcc Fe. 

In the case of systems with a collinear magnetic structure as a ground
state  spin spirals can be
treated as transverse spin fluctuations in the adiabatic approximation.
The energy dispersion of such fluctuations $\Delta E(\vec{q})$ gives
 access to the spin stiffness and exchange coupling constants  of a system
and in this way to the spin excitation spectrum  as well as finite temperature 
magnetism \cite{RJ97, HEPO98, Kueb00}.
 An important feature of   spin spiral
calculations is  that they account 
 for  longitudinal fluctuations  of  the magnetic moment. This leads to more reliable results for    $\Delta E(\vec{q})$  compared to
those obtained using  the  non-self consistent
force-theorem approach.

As was pointed out by Sandratskii and K\"ubler \cite{SK92} the technique
for spin spiral calculations can be used for calculations of 
the static ($\omega=0$) momentum-resolved longitudinal magnetic susceptibility.
Untill now only few  corresponding ab-initio calculations  have been presented
 in the literature.
 In most cases the static
 $q$-dependent magnetic susceptibility was calculated using 
perturbation 
 theory \cite{SPG+00} or  performing
super-cell calculations \cite{Jar86}. 
The spin spiral method, on the other hand,
allows to   perform self-consistent calculations  of  the magnetic susceptibility
avoiding the  super-cell concept \cite{SK92}.

All spin spiral calculations  were
 done so far    using the ASW 
 \cite{SG86, USK94, Kueb00}   or LMTO  \cite{MLSG91,HEPO98}  band structure
 methods. These methods use  Bloch-function basis sets  to represent the solution of  the Kohn-Sham equation and for that reason are restricted to  ordered materials  concerning then application. 
Use of   multiple scattering theory in combination with 
CPA  (Coherent Potential Approximation)
alloy theory, on the other hand,   substantially extends
the variety of materials which can be investigated by
 giving access to  systems without 
chemical order.
Here we present the implementation of spin spiral approach within the
 Korringa-Kohn-Rostoker (KKR) Green's function  band structure method \cite{Ebe00}.
We will show  results of calculations for different systems 
focusing on  disordered alloys.

\section{Theoretical background \label{TB}}

\begin{figure}
\includegraphics[width=45.0mm,angle=0,clip]{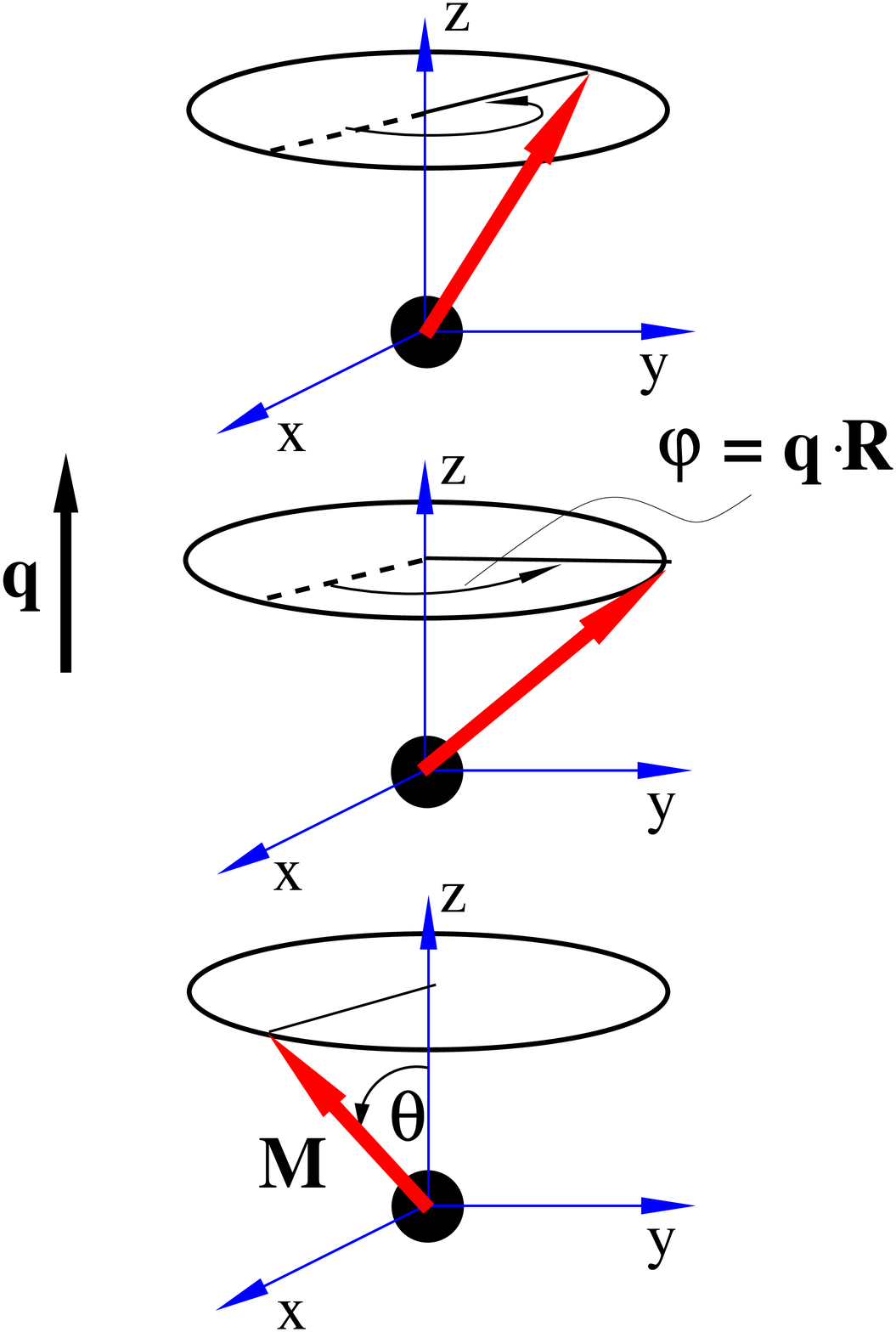}
\caption{\label{fig:SpinSpiral} Geometry of a spin spiral with the wave
  vector $\vec{q}$ along the z-direction. } 
\end{figure}

 When dealing with the electronic structure  of solid state systems
 having a spiral magnetic structure  rotations  can be applied
 independently to the spin and  spatial parts of the electronic wave
 function if spin-orbit coupling (SOC) is neglected.  Using a spin-diagonal 
form of the exchange-correlation
potential  in the local frame of reference  of an atom site, the Kohn-Sham equation 
 for the spinor wavefuction  $\psi(\vec{r})$
can be written in the form:
\begin{equation} 
\left[
-\nabla^2
  \left(
\begin{array}{rr}
 1  & 0 \\
 0  & 1
\end{array}  \right)
 + \sum_{n\nu} U_{n\nu}^\dagger(\theta_\nu,\phi_\nu)
  \left(
\begin{array}{rr}
 V_{n\nu}^+(\vec{r})  & 0 \\
 0  & V_{n\nu}^-(\vec{r})
\end{array}  \right)
 U_{n\nu}(\theta_\nu,\phi_\nu)
\right] \psi(\vec{r}) =  E \psi(\vec{r}) \;.
\label{Eq_KS}
\end{equation}
 Here $\vec{r}_\nu$ denotes
a position of  an atom in a unit cell, $\vec{R}_n$ is a  Bravais lattice vector
  and  
$U_{n\nu}$ is a spin transformation  matrix that connects
 the global
frame of reference  of the crystal to the local frame  of the atom site at $\vec{r}_\nu + \vec{R}_n$
 that has its magnetic moment tilted  away from the global z-direction. 
The transformation $U_{n\nu}$  is characterized by the Euler angles
$\theta_{n,\nu}$ and $\phi_{n,\nu}$  as is is shown in Fig.\
 \ref{fig:SpinSpiral}  for the case of a spin spiral.

As was shown by Sandratskii, considering  a spin spiral structure, 
Eq.\ (\ref{Eq_KS}) can be easily  dealt with using  the properties of spin
space groups (SSG)  \cite{BE66a,BE66b, Her66} 
allowing independent transformations  within the spin and space sub-spaces.
The spin spirals characterized by the wave vector $\vec{q}$ angles $\theta_{\nu}$ and $\phi_{\nu}$  are
represented by  the expression:
\begin{equation*}
\vec{m}_n^{\nu} = m^{\nu}\,
[\mbox{cos}(\vec{q}\cdot\vec{R}_n + \phi_{\nu})\mbox{sin}\theta_{\nu}, 
 \mbox{sin}(\vec{q}\cdot\vec{R}_n + \phi_{\nu})\mbox{sin}\theta_{\nu},
\mbox{cos}\theta_{\nu}]
\label{spiral}
\end {equation*}
defining the spin direction at every site $ (n,\nu)$
of the lattice   
with $m^{\nu}$  the magnitude of  the magnetic moment  on site $\nu$  within the unit cell.
Assuming a collinear alignment of  the
spin density within the atomic cell  at $ (n,\nu)$,    
it's natural to use  a local frame of reference  with its z-axis oriented along $\vec{m}_n^{\nu}  $.
The  corresponding transformation matrices $U_{n\nu}$  occurring in Eq.\  (\ref{Eq_KS}) can be written as a
 product of  two independent rotation  matrices
$U^{\vec{q}}_{n\nu} = U_{n}(\theta_\nu, \phi_\nu, \vec{q}) =
U_{\nu}(\theta_\nu, \phi_\nu)U_{\vec{q}\vec{R}_n}$ where  the matrix
$U_{\vec{q}\vec{R}_n}$  depends only on the translation vector $\vec{R}_n$:\cite{San86,San91}
\begin{eqnarray}
& & U_{n}(\theta_\nu,\phi_\nu,\vec{q}) = \nonumber \\
 &=&
\left(
\begin{array}{rr}
 \cos \frac{\theta_\nu}{2}  & \sin \frac{\theta_\nu}{2} \\
-\sin \frac{\theta_\nu}{2}  & \cos \frac{\theta_\nu}{2}
\end{array} 
\right) 
\left(
\begin{array}{rr}
e^{\frac{i}{2}\phi_\nu }  & 0 \\
0 & e^{-\frac{i}{2}\phi_\nu }
\end{array} 
\right) 
\left(
\begin{array}{rr}
e^{\frac{i}{2}(\vec{q}\cdot \vec{R_n}) }  & 0 \\
0 & e^{-\frac{i}{2}(\vec{q}\cdot \vec{R_n}) }
\end{array}
\right) 
%
 = U_{\nu} \, U_{\vec{q}\vec{R}_n} \;.
\label{spiral_U}
\end{eqnarray}

Instead of  solving  the Kohn-Sham
 Eq.\ (\ref{Eq_KS})  for the eigen functions and values  the electronic structure can be represented in terms of  the corresponding Green's function.
 Within multiple scattering theory the Green's function  is represented in real space by the scattering path operator $\tilde{\underline{\tau}}^{n n'}$  together with the
 regular $Z^{n}_{\Lambda_1}(\vec{r},E)$ and irregular $J^{n}_{\Lambda_1}(\vec{r},E)$ solutions of  the  single-site Kohn-Sham equation  referring to the local frame of reference:
\begin{eqnarray}
G^+(\vec{r},\vec{r}\;',E) & = &
\sum_{\Lambda_1\Lambda_2} 
Z^{n}_{\Lambda_1}(\vec{r},E)
                              {\tau}^{n n'}_{\Lambda_1\Lambda_2}(E)
Z^{n' \times}_{\Lambda_2}(\vec{r}\;',E)
 \nonumber \\
 & & 
-  \sum_{\Lambda_1} \Big[ 
Z^{n}_{\Lambda_1}(\vec{r},E) J^{n \times}_{\Lambda_1}(\vec{r}\;',E)
\Theta(r'-r)  \nonumber 
\\
 & & \qquad\quad 
J^{n}_{\Lambda_1}(\vec{r},E) Z^{n \times}_{\Lambda_1}(\vec{r}\;',E) \Theta(r-r')
\Big] \delta_{nn'} \; .
\end{eqnarray}
  The scattering path operator  is defined by its  equation of motion:
\begin{eqnarray}
{\underline{\tau}}^{n\nu\, n' \nu'} 
 &=&
 {\underline{t}}^{n\nu}\delta_{n\nu\, n' \nu'}
+
{t}^{n\nu} 
 {\sum_{k p}}'
{\underline{G}}^{nq\, m \mu}
 \,
{\underline{\tau}}^{m \mu \, n' \nu'}
\label{tau}\;,
\end{eqnarray}
 where  ${t}^{n\nu} $ and ${\underline{G}}^{nq\, m \mu}$ are the single-site t-matrix and free-electron propagator, respectively, that are all expressed with respect to a common global frame of reference.
Eq.\ (\ref{tau})  has the formal solution
\begin{eqnarray}
 \underline{\underline{{\tau}}}
 &=&
\left[ 
 \underline{\underline{{t}}}^{-1}
- \underline{ \underline{{G}}}
\right]^{-1}
\label{tauf} \; .
\end{eqnarray} 
In Eqs.\ (\ref{tau}) and (\ref{tauf})
 the underline 
indicates  matrices in  the $(l,m)$ representation  while double underline
indicates super-matrices including the site index. In the case of
 a collinear magnetic structure the local and global frames of reference
coincide. This implies that Eq.\ (\ref{tauf}) gives immediately the solution  with respect to the
 local frame of reference.  
For  infinite systems having  a
regular periodic lattice
a solution  to
Eq.\ (\ref{tau})  can be  obtained  by Fourier transformation instead of using the real space expression given in
Eq.\ (\ref{tauf}) . 

For non-collinear magnetic solids with  a periodic lattice structure one
can  solve Eq.\ (\ref{tau}) as for collinear systems but
 using an extended  super-cell. 
The size of  the corresponding
super-cell
is determined by  the period of magnetic structure.
 All  atoms within
the cell are in general  inequivalent and have their own local frame of
reference. Therefore  super-cell calculations can be rather
 time 
consuming  in particular for magnetic structures  having a
 long period.

However, as 
pointed  by various authors \cite{BE66a,BE66b, Her66} 
use of symmetry  allows to simplify the problem substantially.
Spiral magnetic structures  transform according to the
group of generalized translations  that are
characterized by the wave vector
$\vec{q}$ and represented by the matrices $U_{\vec{q}\vec{R}_n}$
(Eq.\ (\ref{spiral_U})).  This implies in particular that the matrices $U_{\vec{q}\vec{R}_n}$  allow to express the single-site t-matrix $ {t}^{n\nu} $ at site $ ({n\nu})$  to that at site
 $ ({0\nu})$. This symmetry
property  allows to write the scattering path operator referring the global frame of
reference as follows:
\begin{eqnarray}
\tau^{n\nu\, n' \nu'} 
 &=&
t^{n\nu} \delta_{n\nu\, n' \nu'}
+
t^{n\nu} {\sum_{m\mu}}' 
G^{n\nu\, m\mu} \tau^{m\mu \, n' \nu'} \nonumber \\
 &=&
 U_{n\nu}^{\vec{q}\,\dagger} \tilde{t}^{n\nu} U_{n\nu}^{\vec{q}} \delta_{n\nu\, n' \nu'}
+
 U_{n\nu}^{\vec{q}\,\dagger} \tilde{t}^{n\nu} U_{n\nu}^{\vec{q}} 
 {\sum_{m\mu}}'
G^{n\nu\, m \mu} \tau^{m\mu \, n' \nu'} 
\label{TAU_R_global} \; .
\end{eqnarray} 
 This allows to find  the scattering path operator and 
 from this
the Green's function in the local frame of reference of
the  each atom, solving the equation
\begin{eqnarray}
\tilde{\underline{\tau}}^{n\nu\, n' \nu'}
&=& U_{n\nu}^{\vec{q}}
\underline{\tau}^{n\nu\, n' \nu'} 
 U_{n'\nu'}^{\vec{q}\,\dagger}\nonumber \\
 &=& 
\tilde{\underline{t}}^{n\nu} \delta_{n\nu\, n' \nu'}
+
\tilde{\underline{t}}^{n\nu} 
 {\sum_{m \mu}}'
U_{n\nu}^{\vec{q}}    \underline{G}^{n\nu\, m \mu}    U_{m\mu}^{\vec{q}\,\dagger}
 \,
U_{m\mu}^{\vec{q}} \underline{\tau}^{m\mu \, n' \nu'} U_{n'\nu'}^{\vec{q}\,\dagger} 
\\
 &=& 
\tilde{\underline{t}}^{\nu} \delta_{n\nu\, n' \nu'}
+
\tilde{\underline{t}}^{\nu} 
 {\sum_{m \mu}}'
\underbrace{
U_{n\nu}^{\vec{q}}    \underline{G}^{n\nu\, m \mu}    U_{m\mu}^{\vec{q}\,\dagger}}_{
\tilde{\underline{G}}^{n\nu\, m\mu} }
 \,
\underbrace{
U_{m\mu}^{\vec{q}} \underline{\tau}^{m\mu \, n' \nu'} U_{n'\nu'}^{\vec{q}\,\dagger} }_{
\tilde{\underline{\tau}}^{m\mu \, n' \nu'} } 
\;, 
\label{TAU_R}
\end{eqnarray} 
where  the tilde indicates matrices which refer to the local frame of reference.  

 In the last  line of Eq.\  (\ref{TAU_R})  use has been made that the single-site t-matrices $\tilde{\underline{t}}^{n\nu} $
 do not depend on the lattice index $ n $ but only on the site index $ \nu $ in the  unit cell.
 As a consequence, the multiple scattering problem can be solved as for the
case of collinear magnetic structures by  Fourier transformation of the equation of motion for the
scattering path operator.
This  leads to its representation in  reciprocal space
 according to:
\begin{eqnarray}
\underline{\underline{\tilde{\tau}}}(\vec{k},E)
 &=&
\left[ 
\underline{\underline{\tilde{t}}}^{-1}(E)
- \underline{\underline{\tilde{G}}}(\vec{k})
\right]^{-1} \; .
\end{eqnarray} 
The structural Green's function  referring to the local
frame of reference can be determined  as follows:
\begin{eqnarray}
\tilde{\underline{G}}^{\nu\nu'}(\vec{k})
 &=&
\frac{1}{N}
\sum_{n n'}
e^{-i\vec{k} \cdot (\vec{R}_{n}-\vec{R}_{n'}) }
\tilde{\underline{G}}^{n\nu\, n'\nu'}
\nonumber \\
 &=&
\frac{1}{N}
\sum_{n n'}
e^{-i\vec{k} \cdot (\vec{R}_{n}-\vec{R}_{n'}) }
 U_{n\nu}^{\vec{q}}
\underline{G}^{n\nu\, n' \nu'}
 U_{n'\nu'}^{\vec{q}\,\dagger}
\nonumber  \\
 &=&
 U_\nu
\left(
\begin{array}{rr}
\underline{G}^{\nu \nu'}(\vec{k}-\frac{1}{2}\vec{q})  & 0 \\
0 & \underline{G}^{\nu \nu'}(\vec{k}+\frac{1}{2}\vec{q})  
\end{array}
\right) 
 U_{\nu'}^{\dagger}
\nonumber \\
 &=&
 U_{\nu} \, \underline{G}^{\nu\nu'}_{\vec{q}}(\vec{k}) \, U_{\nu'}^{\dagger} \;.
\end{eqnarray} 
Here $ \underline{G}^{\nu \nu'}(\vec{k}) $ is a
structural Green's function  for one spin channel represented in  the global
frame of reference.

The charge distribution  within  the central unit cell $n=0$ is determined by 
the cell-diagonal scattering path operator
 $\tilde{\underline{\underline{\tau}}}^{00}$
 which can be found by 
  the Brillouin zone integral 
\begin{eqnarray}
\tilde{\underline{\underline{\tau}}}^{00}
 &=&
\frac{1}{\Omega_{BZ}}
\int_{\Omega_{BZ}} d^3k
\left[ \tilde{ \underline{\underline{t}}}^{-1} - \tilde{\underline{\underline{G}}}(\vec{k}) 
\right]^{-1}  \nonumber 
\\
 &=&
 U_{0}^{\dagger}
\frac{1}{\Omega_{BZ}}
\int_{\Omega_{BZ}} d^3k
\left[ { \underline{\underline{t}}}^{-1} - 
 \underline{\underline{G}}_{\vec{q}}(\vec{k})
\right]^{-1} 
 U_{0}
\nonumber \\
 &=&
 U_{0}^{\dagger}
\underline{\underline{\tau}}^{00}
 U_{0} \;,    
\end{eqnarray} 
where  $U_{0}$ is the  transformation
 matrix diagonalising the potentials as well as 
$t$-matrices with respect to spin  within the  central unit cell.

To perform  calculations for
disordered alloys the CPA (Coherent Potential Approximation) alloy
theory \cite{MSS90,But85} is used. In  the case  of a spin spiral system the CPA medium is
represented in  the
global frame of reference by the
effective single-site scattering matrix $\underline{t}^C$ and the scattering path operator  obtained from the expression:
\begin{equation}
  \label{eq:tauc}
  \underline{\underline{\tau}}^{00 \rm, C}(E) = \frac{1}{\Omega_{\rm BZ}} \int_{\Omega_{\rm BZ}}
  d^3k \left [ (\underline{\underline{t}}^{\rm C}(E))^{-1} - \underline{\underline{G}}(\vec{k},E)\right ]
  ^{-1} 
  \; .
\end{equation}
The  corresponding element  projected scattering path operators are  obtained from  these via:
\begin{equation}
\begin{aligned}
  \label{eq:taualpha}
  \underline{\underline{\tau}}^{00,\alpha} &\,=\,\underline{\underline{\tau}}^{00,\rm C} \,
  \underline{\underline{D}}^{\alpha}\,, 
\end{aligned}
\end{equation}
with 
\begin{equation}
\begin{aligned}
  \label{eq:dmatrix}
  \underline{\underline{D}}^{\alpha} & \,=\,  \left [ \underline {\underline{1}} + 
                           \left [
               (\underline{\underline{t}}^{\alpha})^{-1} - (\underline{\underline{t}}^{\rm C})^{-1}
                           \right ] \underline{\underline{\tau}}^{00,\rm C}
                           \right ] ^{-1} \,.
\end{aligned}
\end{equation}

The approach developed for calculations of non-collinear spin spiral
structures can be used 
for  investigations  on the 
longitudinal magnetic susceptibility as a
function of  the wave vector $\vec{q}$ \cite{SK92}. 
This  approach allows to avoid the use of perturbation theory and can be
 applied  to magnetic  as well as non-magnetic systems.
 In the following we focus on materials  in  their non-magnetic    state which   may exhibit  
 paramagnetism (Ag$_x$Pt$_{1-x}$), ferromagnetism (Ni) or antiferromagnetism (Cr) in
their ground state. 
For  this purpose we specify a spiral external magnetic field to be
perpendicular to the direction of  the wave vector $\vec{q}$ (i.e.\ $\theta = 90^o$): 
\begin{equation*}
\vec{h}(\vec{r}) = h_0
[\mbox{cos}(\vec{q}\cdot\vec{R}_n),\mbox{sin}(\vec{q}\cdot\vec{R}_n),0]
\label{spiral}\; .
\end {equation*}
In this
case is  the potential energy term in  the Kohn-Sham equation (see Eq.\ (\ref{Eq_KS})) 
 is given by:
\begin{equation} 
\left[
 \sum_{n\nu} U_{n\nu}^\dagger(\theta_\nu,\phi_\nu)
  \left(
\begin{array}{cc}
 V_{n\nu}(\vec{r}) - B_{n}^{ext} - \Delta B_{n}^{xc}(\vec{r}) & 0 \\
 0  & V_{n\nu}(\vec{r}) + B_{n}^{ext} + \Delta B_{n}^{xc}(\vec{r})
\end{array}  \right)
 U_{n\nu}(\theta_\nu,\phi_\nu)
\right] \;.
\label{Eq_KS_susc}
\end{equation}
A self-consistent
calculation  based on
Eq.\ (\ref{Eq_KS_susc}) gives the spin magnetic moment induced by the external magnetic
field. The $\vec{q}$-dependent external magnetic field should be taken
small enough to be considered  as a perturbation. 
In this case, assuming a linear response to be the leading term of the response function
 the  corresponding magnetic susceptibility can be derived from the expression 
\begin{eqnarray}
\chi(\vec{q})
 &=&
\frac{m_{ind}(\vec{q})}{h_0}
\label{chi}  \; .
\end{eqnarray} 
Suppressing  the spin-dependent part of the exchange-correlation
potential ($\Delta B_{n}^{xc}(\vec{r}) = 0$), one 
can calculate the unenhanced spin susceptibility $\chi^0(\vec{q})$.  
Otherwise,  Eq.\ (\ref{chi}) gives  the 
enhanced longitudinal magnetic susceptibility $\chi(\vec{q})$, 
represented in linear response theory for uniform system by the expression
\begin{eqnarray}
\chi(\vec{q})
 &=&
\frac{\chi^0(\vec{q})}{1 - I(\vec{q})\chi^0(\vec{q})}
\label{chi_q}  \;, 
\end{eqnarray} 
with $I(\vec{q})$ the exchange integral responsible for the enhancement
of the magnetic susceptibility (see, for example,
\cite{Kueb00,Den01}). 
 In case of a paramagnetic ground state the magnetic 
susceptibility $\chi(\vec{q})$ is positive  for all values of
$\vec{q}$. For  other cases
the denominator in the Eq.\ (\ref{chi_q})  may become zero  or even negative.  This  singular behavior of the susceptibility obviously
 indicates an
instability of the paramagnetic state  towards a transition to spontaneous formation of ferro- or anti-ferronagnetic  order.

\section{ Results }
\subsection{\label{sec:spinspiral} Spin spiral structure in alloys}

In the following  several  applications of the scheme introduced  above
  are presented  that focus  on disordered alloys 
to demonstrate the  flexibility of the multiple scattering formalism  when dealing with
spin spiral  systems.  Corresponding calculations have been performed for
 alloys having fcc (Fe$_x$Ni$_{1-x}$ and Fe$_x$Mn$_{1-x}$) and bcc
(Fe$_x$Co$_{1-x}$) lattice structure.

 For all calculations  the angle $\theta$  has been chosen to
 be $90^o$.  For this spin geometry, the spin spiral with
$\vec{q}=\frac{\pi}{a}(0,0,1)$ corresponds to  a spin configuration
where the first neighbor atoms  in  the (001) direction have an  anti-parallel
(AFM) spin alignment, while    $\vec{q}= \vec{0}$  implies a parallel (FM) orientation.

\subsubsection{ The disordered alloy system Fe$_x$Ni$_{1-x}$ }

\begin{figure}
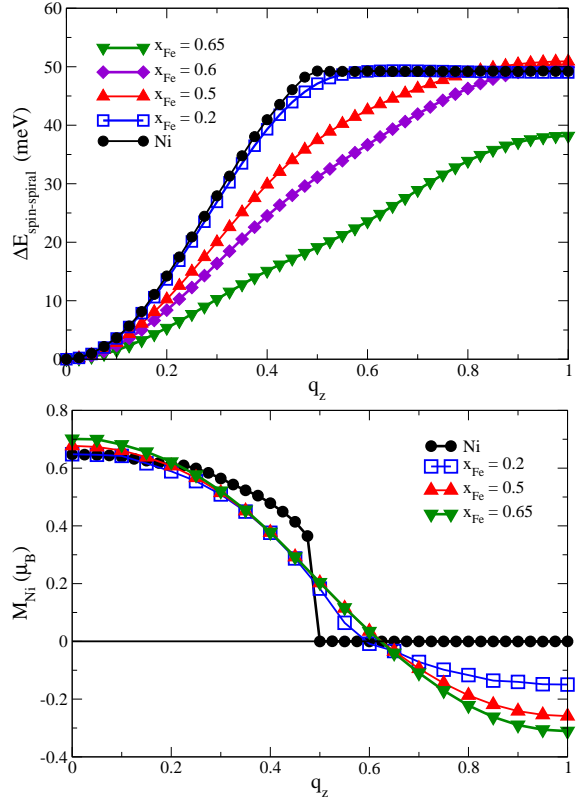

\includegraphics[width=75.0mm,angle=0,clip]{FeNi_spinspiral_cmp_2.eps} \\
\includegraphics[width=75.0mm,angle=0,clip]{FeNi_spinspiral_Ni_MMOM_2_cmp.eps}
\caption{\label{fig:FeNi_2} a) The energy of spin spiral magnetic structure
 in Fe$_x$Ni$_{1-x}$ alloys and b) local
 magnetic moments on Ni atoms  as a function
 of the wave vector $\vec{q} = \frac{\pi}{a}(0,0,q_z)$. } 
\end{figure}

 Fig.\ \ref{fig:FeNi_2}a shows the energy of 
the disordered Fe$_x$Ni$_{1-x}$ alloy system   with a spin spiral structure as
a  function of  the wave  vector $\vec{q}$.
 For all  concentrations the minimum of the 
 energy  is found for $\vec{q}= \vec{0}$,  implying 
 that the ferromagnetic structure  is  more stable configuration
 than non-collinear
  structures characterized by wave vectors along the (001) direction.   

\begin{figure}
\includegraphics[width=75.0mm,angle=0,clip]{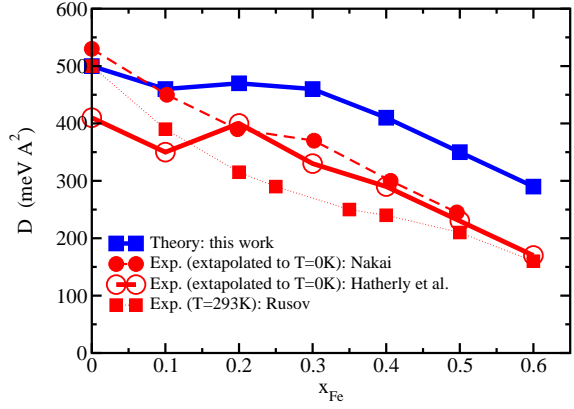}
\caption{\label{fig:FeNi_3} Spin stiffness constant of
 Fe$_x$Ni$_{1-x}$ alloys   as a function of the concentration in 
comparison with experiment: 
Nakai \cite{Nak83} (full circles),
  Hatherly et al.\ \cite{HHL+64} (open circles), 
Rusov \cite{Rus76} (full squares).} 
\end{figure}

 As can be seen from 
 Fig.\ \ref{fig:FeNi_2}b, the local magnetic moment of Ni  in pure Ni drops down
to $m =0$ at the wave vector $\vec{q}_c \approx \frac{\pi}{a}(0,0,0.5)$
and the 
system becomes paramagnetic. In terms of the Stoner theory of ferromagnetism
(see, e.g.\  Ref.\ \onlinecite{Kueb00})
this means that the criterium for the instability of  the paramagnetic state  
is  satisfied only  for small  wave vectors, while above $\vec{q}_c$
the paramagnetic  (PM) state should be  the most stable state of  the system. 
The criterion  for the instability of the PM state will be discussed
below in more detail.

   Adding only small amounts of  Fe  to Ni leads  obviously to a
nonzero magnetic moment per unit cell at all values of wave vector $\vec{q}$.
This is caused by the large  magnetic moment  of  Fe which 
depends  only slightly on the wave vector.
 Fig.\ \ref{fig:FeNi_2}b  shows that the Ni magnetic
moment in contrast to  that of Fe,  varies rather rapidly  with increasing
 wave vector  and changes  sign  at  $\vec{q} \approx
\frac{\pi}{a}(0,0,0.6)$.
 This means that  in the vicinity of  the ground state of the alloys
($\vec{q}= \vec{0}$) the magnetic moments of  Fe and Ni  atoms
prefer to have parallel  alignment, while close to
$\vec{q}=\frac{\pi}{a}(0,0,1)$ (AFM structure along (0,0,1) direction)
 the more favorable orientation of the Fe and Ni  moments is anti-parallel (AP).
Nevertheless, even for small Fe concentrations, the total
magnetic moment  is determined by  the dominating moment of Fe.
 As a result, the
alloy system exhibits  effectively a  ferromagnetic behavior	
 for all wave vectors, as one can see in Fig.  \ref{fig:FeNi_2}.

The energy difference between the spin spiral
states with $\vec{q}= \vec{0}$ and $\vec{q} = \frac{\pi}{a}(0,0,1)$ 
 remains almost unchanged up to the Fe concentration $x_{Fe} \approx 0.6$,
and changes nearly by 20 \%  when approaching  $x_{Fe} \approx 0.65$.
  On the other hand,
the spin-stiffness constant  deduced from the energy
dispersion curves
decreases continuously with the increase
of Fe content as can be seen from Fig.\  \ref{fig:FeNi_3}.
    This figure also shows that the calculations reproduce the
 available experimental data  for the 
spin-stiffness  constant fairly well, although they seem to be slightly to high.
 This difference can be partially attributed
to the conditions of the experiment  as e.g.\ polycrystallinity
of the samples and a finite temperature.

\subsubsection{The disordered alloy  Fe$_{0.5}$Co$_{0.5}$ }

\begin{figure}
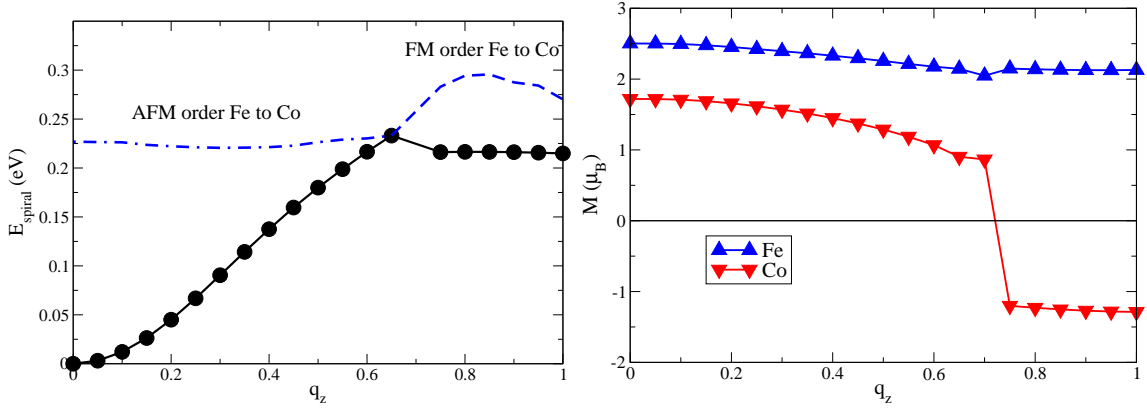

\includegraphics[width=75.0mm,angle=0,clip]{FeCo_CMP_E_spiral_2.eps} 
\includegraphics[width=75.0mm,angle=0,clip]{FeCo_CMP_M_spiral_1.eps}
\caption{\label{fig:FeCo_1}a) the energy of spin spiral magnetic structure
 in Fe$_{0.5}$Co$_{0.5}$  calculated for the wave vector $\vec{q} =
 \frac{\pi}{a}(0,0,q_z)$  along [001] direction; b) local
 magnetic moments on Fe and Co atoms separately, as a function of wave
 vector of spin spirals.   } 
\end{figure}

The          change of sign of the magnetic moment observed for
 Fe$_x$Ni$_{1-x}$ alloys for one
of the alloy components upon variation of the wave vector becomes even more 
pronounced in bcc Fe$_{0.5}$Co$_{0.5}$  and fcc Fe$_{0.5}$Mn$_{0.5}$ alloys.
Disordered bcc Fe$_{0.5}$Co$_{0.5}$  has  a ferromagnetic ground state.
The spin spiral energy  shown in Fig.\ \ref{fig:FeCo_1}a increases
  with  wave vector  confirming  the stability of the FM state.
As  can  be seen from Fig.\ \ref{fig:FeCo_1}b, around
$\vec{q}= \vec{0}$ the  individual Fe and  Co  moments are aligned parallel with
respect to each other. 
However, after crossing 
$\vec{q}_c \approx \frac{\pi}{a}(0,0,0.7)$, 
the total magnetic moment jumps from $m = 1.46
\mu_B$ to $m = 0.47 \mu_B$  due to  a change of the sign of  the Co magnetic
moment  with respect to that of the dominating Fe moment. 
As  Fig.\ \ref{fig:FeCo_1}b   shows, the dispersion of  the spin spiral energy
 for the anti-parallel configuration gets very weak up to $\vec{q} \approx
\frac{\pi}{a}(0,0,1)$. 
To  estimate the energy of the spin spirals  for the non-equilibrium  
configurations, i.e.\ anti-parallel for $\vec{q}_c   < \frac{\pi}{a}(0,0,0.7)$
and  parallel for $\vec{q}_c  >  \frac{\pi}{a}(0,0,0.7)$,   respectively,
   frozen potential calculations 
have been performed. The corresponding 
results are  represented in Fig.\ \ref{fig:FeCo_1}b by dashed and
dashed-dotted lines. Obviously, these results augment  the two stable
  branches fairly will.

\subsubsection{The disordered alloy   Fe$_{0.5}$Mn$_{0.5}$ }

\begin{figure}
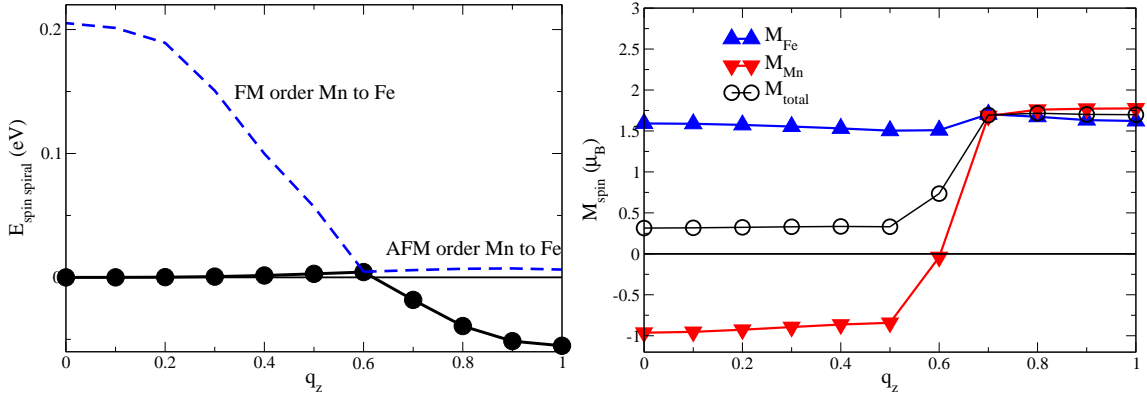

\includegraphics[width=75.0mm,angle=0,clip]{FeMn_E_spinspiral_CMP.eps}
\includegraphics[width=75.0mm,angle=0,clip]{FeMn_Mspin_spinspiral_CMP_main.eps}
\caption{\label{fig:FeMn_1} a) The energy of spin spiral magnetic structure
 in Fe$_x$Mn$_{1-x}$ alloys  calculated for the wave vector $\vec{q} =
 \frac{\pi}{a}(0,0,q_z)$ along [001] direction; b) local
 magnetic moments on Fe and Mn atoms separately, as a function of wave
 vector of spin spirals.   } 
\end{figure}

Fig.\ \ref{fig:FeMn_1} shows the results of spin spiral calculations for
Fe$_{0.5}$Mn$_{0.5}$ having a non-collinear magnetic structure as a ground
state \cite{JPS88,SBS+99,EI71}.  
As is seen in the energy dispersion curve, Fig.\ \ref{fig:FeMn_1}a, the
system exhibits the behavior of a FM system for wave vectors $\vec{q}$
up to  $\vec{q}_c \approx \frac{\pi}{a}(0,0,0.6)$. 
In this wave vector region 
the alloy has  a small   average magnetic moment
formed by two anti-parallel aligned magnetic moments of Fe and Mn   occupying randomly the site 
(see Fig. \ref{fig:FeMn_1}b). 

At   $\vec{q}_c \approx \frac{\pi}{a}(0,0,0.6)$ the energy of  a
spin spiral reaches  its maximum and the 
following increase of the wave vector is accompanied by   a decrease in energy and an increase of the  average magnetic moment. At $\vec{q} = (0,0,\frac{\pi}{a})$ the spin spiral magnetic 
structure reaches  its energy minimum, which is about 50 meV lower
than the energy of  the FM state, with  a parallel alignment of the
magnetic moments of  the alloy components.

Similar to Fe$_{0.5}$Co$_{0.5}$, 
these two minima of the energy  -- around $\vec{q} =
\vec{0}$ and around  $\vec{q} = (0,0,\frac{\pi}{a})$ -- 
 are formed by two crossing
branches of  the spin spiral  dispersion relation: 
one corresponds to  an anti-parallel alignment of  the
Fe and Mn
magnetic moments (around the $\vec{q}= \vec{0}$) and another  to
their parallel alignment (around $\vec{q} =
(0,0,\frac{\pi}{a})$ ), which have a crossing point at $\vec{q} \approx
\frac{\pi}{a}(0,0,0.6)$.

Thus, from the analysis of the energetics of  the spin spiral structures in
Fe$_{0.5}$Mn$_{0.5}$, one can conclude that the system  has in its magnetic
ground 
state  an anti-parallel alignment of  the magnetic moments of  first
neighbors, no matter  whether the neighboring atoms are Fe or Mn.

\subsection{ Spin susceptibility }
\label{sec:suscept}

In the present section we will discuss another application 
of the technique  presented above. 
As was shown by Sandratskii and K\"ubler \cite{SK92}, 
spin spiral calculations can also be used 
to determine the longitudinal magnetic susceptibility 
$\chi$, both for magnetic and non-magnetic systems, as a function of
 the
wave vector $\vec{q}$. 
This approach allows  in particular to avoid the use of perturbation theory.
  Adding a Zeeman term to the Hamiltonian corresponding to a
small external helical magnetic field 
allows  to obtain the magnetic susceptibility from the
the induced magnetic
moments.  For the present calculations a Zeeman splitting
 $h_0 = 1$ meV  has been used.

 The present work  deals with non-magnetic systems, which   have  either
 a paramagnetic (AgPt), a ferromagnetic (Ni) or an  anti/ferromagnetic (Cr) 
 ground state. 
Dealing with  magnetic systems being in an  imposed paramagnetic state,
 their magnetic susceptibility  gives information on an 
instability with respect to magnetic ordering.

\subsubsection{ The  paramagnetic disordered alloy  Ag$_x$Pt$_{1-x}$ }

\begin{figure}
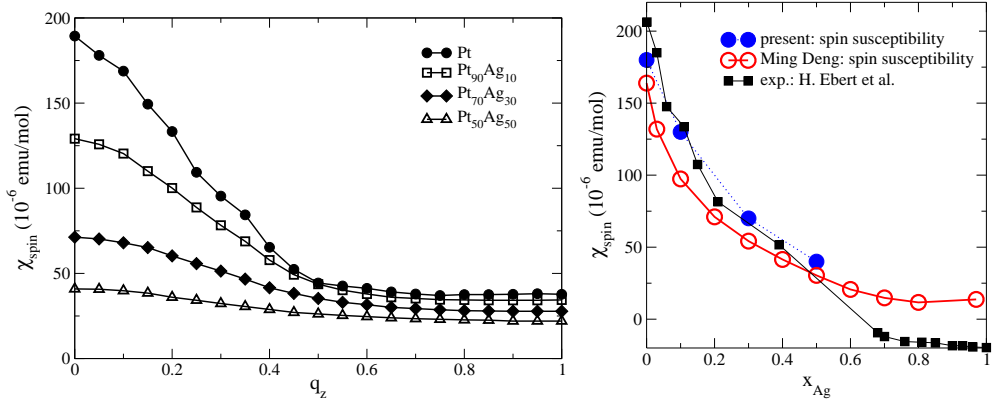

\includegraphics[width=75.0mm,angle=0,clip]{AgPt_suscept_vs_q_lmax3.eps}
\includegraphics[width=55.0mm,angle=0,clip]{CMP_chispin_AgPt.eps}
\caption{\label{fig:AgPt_susc} a) Wave-vector ($\vec{q} =
  \frac{\pi}{a}(0,0,q_z)$) dependent magnetic 
  susceptibility of the paramagnetic disordered  Ag$_x$Pt$_{1-x}$
for various  concentrations. 
b)  Comparison of the present results for  the susceptibility  for
  $q=0$ with the results of Deng et al.\ \cite{Den01} 
obtained   via linear response theory.} 
\end{figure}

Fig.\ \ref{fig:AgPt_susc} shows the magnetic susceptibility
of paramagnetic    Ag$_x$Pt$_{1-x}$  alloys
  as a function of  the wave vector $\vec{q}$\; for various  concentrations. 
The spin susceptibility of the alloys presented in 
Fig.\ \ref{fig:AgPt_susc} is  composed by contributions from
 both components  according to
 $\chi(\vec{q}) = x\chi_{Ag}(\vec{q},x) +
(1-x)\chi_{Pt}(\vec{q},x)$. 
For all concentrations, the increase of the wave vector for 
helical magnetic field is accompanied by a decrease of the response
functions, as it  is usually found  for paramagnetic systems. The main contribution
to  the spin 
susceptibility  stems from the Pt atoms.  As can be seen, increasing the Ag
 content leads to a decrease of  the magnetic susceptibility  for all values of wave
vector.

The present results for $\vec{q} = \vec{0}$ are 
compared with the   total magnetic susceptibility obtained 
 via  fully relativistic linear response  calculations \cite{Den01}. 
As one can see,  the
agreement of results obtained by the two rather different theoretical 
approaches  is rather good. One reason for the observed deviations is
the use of a finite value for the external magnetic field in the present calculations
giving the  magnetic susceptibility from the induced magnetic moment within the
self-consistent calculations. Another reason is 
the neglect of spin-orbit coupling within the present 
calculations that usually reduces the spin susceptibility.
 Nevertheless, both approaches lead obviously to coherent 
results that are in rather satisfying agreement
 with experimental results \cite{EAV84}  (full squares in Fig.\
   \ref{fig:AgPt_susc}b). Note however, that experimental results
   represent the total magnetic susceptibility including also the
   orbital contribution.

\subsubsection{ Pure ferromagnetic  fcc Ni }

\begin{figure}
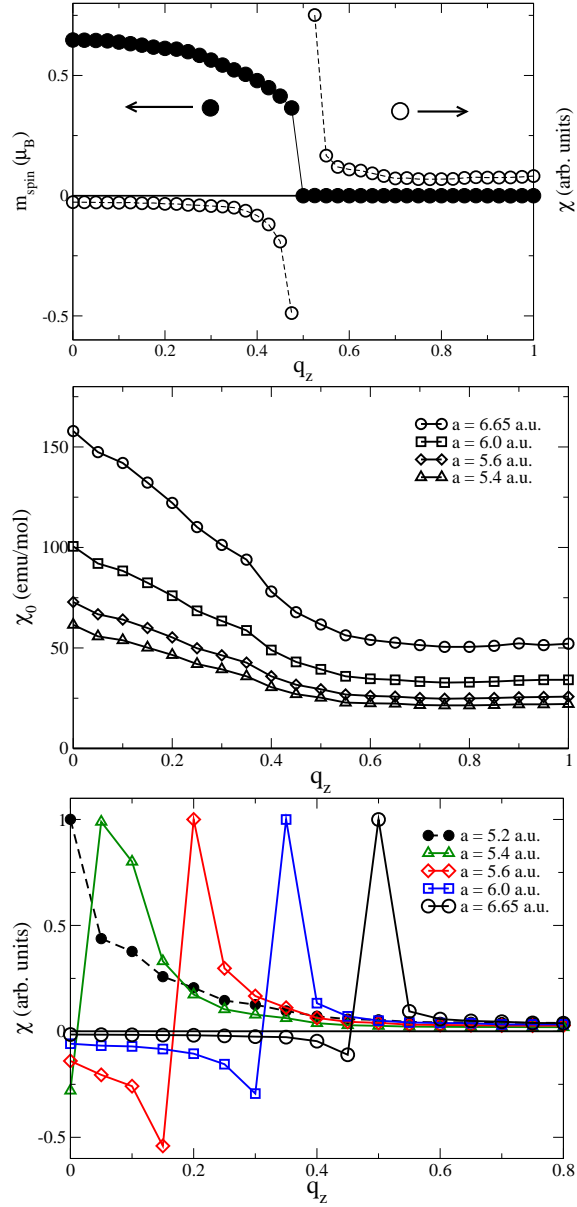

\includegraphics[width=75.0mm,angle=0,clip]{CMP_Ni_PM_susc_Magn.eps} \\
\includegraphics[width=75.0mm,angle=0,clip]{CMP_Ni_PM_unenh_susc_3.eps} \\
\includegraphics[width=75.0mm,angle=0,clip]{CMP_Ni_PM_susc_vs_ALAT_2.eps}
\caption{\label{fig:Ni_susc}  Wave-vector dependent spin susceptibility of paramagnetic Ni having lattice parameter $a = 6.65$ a.u. together with local Ni magnetic moment as a function of wave-vector characterizing non-collinear spiral magnetic structure (a). The wave-vector dependent unenhanced (b) and enhanced (c) magnetic susceptibilities for paramagnetic Ni calculated for different lattice parameters. } 
\end{figure}

The calculations  performed for  ferromagnetic Ni in  a paramagnetic state show a 
behavior for the magnetic susceptibility as a function of  the wave vector  that is rather different  from that of systems 
 with a paramagnetic ground state as for example 
Ag$_x$Pt$_{1-x}$  alloys) (see
Fig.\ \ref{fig:Ni_susc}).
 The paramagnetic state of Ni was simulated
using the disordered local moment (DLM) \cite{GPS+85} method assuming
equal  concentration  for 
atoms with  opposite  orientation of their  magnetic moments. 
The  magnetically disordered state of Ni is characterized by  a  vanishing local
magnetic moment and therefore the DLM method allows us to force the
local magnetic moment to be zero.                     
 Fig.\ \ref{fig:Ni_susc}a shows the results obtained for Ni with the experimental
 lattice parameter $a = 6.65$ a.u. 
 At small values of the wave vector $\vec{q}$ the magnetic susceptibility
 is negative  indicating an
 instability of the paramagnetic state. 
 This is a result of the high density of states (DOS) of  the 3d-electrons
 leading to a large value of  the
unenhanced magnetic susceptibility
 $\chi^0$.    Accordingly, for   small $\vec{q}$-vectors
 the Stoner condition for a
 magnetic instability  $I(\vec{q})\chi^0(\vec{q}) > 1$ (Eq.\ \ref{chi_q})
 (see, e.g., \cite{Mor85,  Mohn03})
 is fulfilled.

As one can see in  Fig.\ \ref{fig:Ni_susc},
at the wave vector $\vec{q} \approx \frac{\pi}{a}(0,0,0.55)$
(for which the denominator in Eq.\ (\ref {chi_q}) comes to 0)
 the
magnetic susceptibility becomes singular and the following increase of
$\vec{q}$ results in  a change of  sign  for the susceptibility  from negative to
positive  leading to the stability of the paramagnetic state. 

 Fig.\ \ref{fig:Ni_susc}a shows also  the Ni magnetic moment as a function of the wave vector 
$\vec{q}$ of spin spiral. As one can see,   the  magnitude of the
 goes down upon increase of $\vec{q}$ 
reaching $m = 0$ at  the critical value of the wave vector $\vec{q}_c$.

As is shown in Fig.\ \ref{fig:Ni_susc}b, a decrease of the lattice
parameter leads to a decrease of the unenhanced susceptibility
$\chi(\vec{q})$ due to the broadening of the energy bands of the
3d-states. This results in a decrease  of the 
critical wave vectors $\vec{q}_c$ until a
 lattice parameter is reached for  which $\vec{q}_c = 0$.
For smaller lattice parameters the ground state of Ni is the PM state.

\subsubsection{ Pure antiferromagnetic bcc Cr }

\begin{figure}
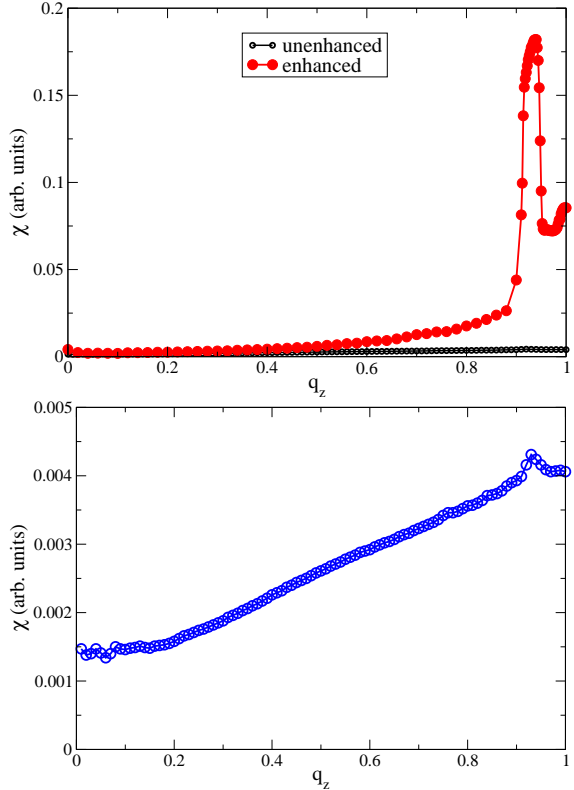


\includegraphics[width=75.0mm,angle=0,clip]{Cr_sus_alat_5.2_vs_5.4.eps}\\
\includegraphics[width=75.0mm,angle=0,clip]{Cr_sus_unenh_alat_5.4.eps}
\caption{\label{fig:Cr_susc} Wave-vector dependent enhanced (a) and unenhanced (b) spin susceptibilities of paramagnetic Cr. For comparison, the unenhanced susceptibility is plotted also at the panel (a).} 
\end{figure}

Results for the non-magnetic state of Cr having the AFM structure as a
ground state are shown in Fig.\ \ref{fig:Cr_susc}. Note that the
antiferromagnetic order of Cr on the one side is 
a result of nearly-half filling of the d-band
\cite{Mor85}  (similar
to Mn), that should result in a commensurate AFM 
structure. However, Cr exhibits also an instability with respect to an
incommensurate spin-density wave (SDW) with the wave vector $\vec{q}
\approx \frac{\pi}{a} (0,0,0.95)$, which is a result of the Fermi
surface nesting. This leads 
to a singularity of the magnetic susceptibility at $\vec{q}
\approx \frac{\pi}{a} (0,0,0.95)$ of paramagnetic Cr.
This SDW instability in Cr and the corresponding behavior of
the momentum dependent
magnetic susceptibility was discussed in the
literature by several authors \cite{SPG+99,CY93,Faw88}. 

Our present results demonstrate that the calculation
of the momentum-resolved magnetic susceptibility properly reproduce its
$\vec{q}$ dependent features for Cr.
The calculations have been performed for a 
 lattice parameter $a = 5.4$ a.u.\ which is slightly
 smaller than
the experimental one ($a \approx 5.44$ a.u.).
 At this lattice parameter
the PM state was found to be more stable than the AFM state. This allows
us to observe the behavior of $\chi(\vec{q})$ 
 due to the Fermi
surface nesting  avoiding the influence of other singularities connected to the instability around
$\vec{q} = \pi/a(0,0,1)$ with respect to the AFM state.   

Fig.\ \ref{fig:Cr_susc}b shows a monotonous increase of the unenhanced
susceptibilities with increasing
 wave vector $\vec{q}$ 
reaching its maximum at $\vec{q} \approx \pi/a(0,0,0.92)$.
 The enhanced
susceptibility, also increasing with 
 wave vector $\vec{q}$, has a
drastic increase at $\vec{q} \approx \pi/a(0,0,0.92)$ due the
enhancement factor (Eq.\ (\ref{chi_q})), which is associated with a
singularity caused by the Fermi surface nesting 
 mentioned above. 

Here, we do not discuss  the $q$ dependence
 of the exchange
integral $I(\vec{q})$ as this was done in detail by Sandratskii and
K\"ubler. Nevertheless, we would like to 
stress that this feature is taken into account within the self-consistent
calculations for every wave
vector. In fact this is essential for the analysis of the
stability of the
paramagnetic state.

\section{ Conclusion }

A theoretical approach for 
  electronic structure calculations on systems with spiral magnetic
  structures within the KKR Green's function formalism
has been presented.
As has been demonstrated, by making use of symmetry, the
scattering path operator can be obtained by
solving the corresponding equation of motion in the
reciprocal space. Compared to the case  of collinear magnetic structure only the structural Green's function to be used involves
the wave vector of the spin spiral.
As the KKR-formalism combined with the CPA  allows to deal with chemically disordered
materials, corresponding spin spiral investigations on various disordered alloys could be performed. In particular the energy of spin spirals  
and the behavior of the magnetic moments of the alloy components was analyzed.
In addition is was shown that
the approach presented
can be efficiently used for the calculation
of the momentum resolved longitudinal magnetic susceptibilities of pure
materials as well as of disordered alloys.

\section{Acknowledgement}

This work was supported by the DFG within the
project Eb 154/20 "Spin polarisation in Heusler alloy based spintronics
systems probed by SPINAXPES".


\end{document}